\documentclass{aa}
\begin{document}
\include{epsf}
   \thesaurus{	11.03.1; 
		12.12.1; 
	        13.25.2} 

   \title{Detection of Filamentary X-Ray Structure in the Core of the Shapley Supercluster}
   \markboth{Filamentary X-Ray Structure in the Shapley Supercluster Core}
{Filamentary X-Ray Structure in the Shapley Supercluster Core}

   \subtitle{}

   \author{A. Kull\inst{3,2,1} and H. B\"ohringer\inst{1}}

   \offprints{kull@cita.utoronto.ca}

   \institute{\inst{1} Max-Planck-Institut f\"ur Extraterrestrische Physik,
              D-85740 Garching, Germany\\
	      \inst{2} Institute for Astronomy,
	      2680 Woodlawn Drive,
              Honolulu, Hawaii 96822, USA\\
	      \inst{3} Present address: Canadian Institute 
  	      for Theoretical Astrophysics, 60 St. George Street, 
	      Toronto, M5S 1A7, Canada}
   \date{Received ... X, XXXX; accepted ... X, XXXX}

   \maketitle

   \begin{abstract}
We report on X-ray observations of the core of the Shapley Supercluster.
Combining data from pointed observations of the ROSAT PSPC detector and 
data from the ROSAT All-Sky Survey, the observed region covers an area of 
$6^\circ \times 3^\circ$. It contains the central clusters A3562, A3558 
and A3556. 

We find clear evidence for X-ray emission connecting the three clusters. 
This confirms the existence of a filamentary, physical structure embedding
the three clusters A3562, A3558 and A3556. We also find evidence for faint emission
westwards of A3556. In total, the extension of the filamentary X-ray emission of 
the core of the Shapley Supercluster amounts up to at least 
$\sim 17.5 h_{50}^{-1}$ Mpc. The total luminosity in the 0.1-2.4 keV energy 
band is $\sim 16\times10^{44} h_{50}^{-2}$ erg s$^{-1}$.

\end{abstract}

\section{Introduction}
The Shapley Supercluster (SSC) has for long been known as a large 
region of a high overdensity of galaxies (Shapley 1930). Located in
in the direction of Hydra-Centaurus at $z \sim 0.046$ it
is one of the densest large scale concentrations
of matter (Fabian 1991) in the Universe. If clusters of galaxies
are taken as mass tracers (Scaramella et al. 1989), the Shapley concentration
accounts for at least 10-20\% of the acceleration of the Local Group
towards the Great Attractor (Lynden-Bell et al. 1988, 
Scaramella et al. 1991, Drinkwater et al. 1998).

Optical (Vettolani et al. 1990, Raychaudhury et al. 1991, Bardelli et al. 1994,
Quintana et al. 1997) and X-ray observations (Raychaudhury et al. 1991, 
Day et al. 1991, Breen et al. 1994, Bardelli et al. 1996, Ettori et al. 1997) 
reveal that the SSC has a remarkable structure  
dominated by a central high-density core. Centered on A3558 (Shapley 8) 
this central region
of the SSC includes five Abell Clusters (A3558, A3556, A3559, 
A3560, A3562) and two rich groups of galaxies (SC1327-312 and SC1329-313). This 
corresponds to an overdensity of $\sim 400$ with respect to the average density 
of ACO clusters (Vettolani et al. 1990).
 
A comprehensive study of the mass distribution in the SSC
inferred from X-ray observations has been reported by Ettori et al. (1997).
Their analysis was based on a mosaic of ROSAT PSCP and Einstein 
Observatory IPC data enclosed in a sky area of $15^\circ \times 20^\circ$ and
centered on A3558. The individual mass estimates for 14 clusters and 
two groups of galaxies amount to an estimated total mass of the SSC 
of $\sim 4 \times 10^{16}$M$_\odot$. Assuming a CDM-like power spectrum, 
Ettori et al. (1997) find that the core of Shapley is an overdensity on the level of 
$3.5\sigma$ ($\Omega=0.3$) to $3.7\sigma$ ($\Omega=1$).

In this letter we report on the X-ray observation of the core of the SSC 
by ROSAT combining pointed observations and data from the ROSAT All-Sky Survey. 
The region of our study is centered at $\alpha_{2000}: 13^{h}28^{min}$ ; 
$\delta_{2000}: -31.5^\circ$ and covers a field of $6^\circ \times 3^\circ$. It
contains the central clusters A3562, A3558 and A3556 and the two rich groups of galaxies 
SC1327-312 and SC1329-313. This part of the core of the SSC is 
of particular interest since there are strong indications (Bardelli et al. 1994) 
that it forms a single elongated structure on the scale of $\sim 15 h_{50}^{-1}$ Mpc
at a redshift of $\sim 0.0482$. 
Our analysis shows that the core of SSC is traced by an elongated X-ray emission
on similar scale. This confirms the existence of a physically connected, filamentary 
structure in the core of Shapley.

\section{X-Ray Observation and Analysis}
While ROSAT PSPC pointed data is limited by the sky coverage, ROSAT 
All-Sky Survey (RASS) (Tr\"umper 1993, Voges et al. 1996) data offers a virtually unlimited 
field of view. On the other hand, RASS data is limited by the short exposure time
(i.e. $\sim 320$ sec vs. typical $10^4$ sec of PSPC pointings). As only a part 
of the core of the SSC is covered by pointed observations, a completion of this area
by RASS data offers a possibility to combine both advantages.

\begin{figure*}
\centering\leavevmode{\epsfxsize=15.0cm\epsffile{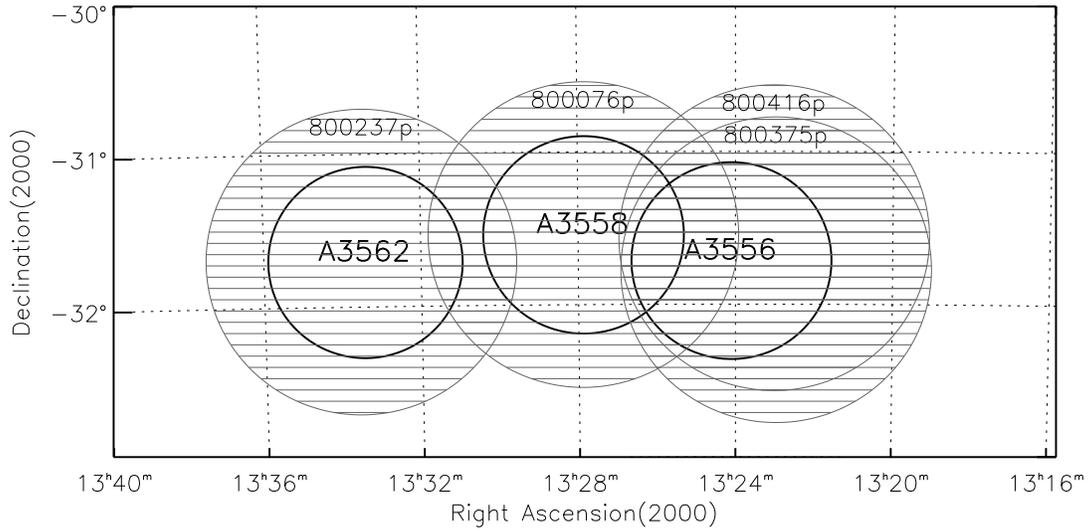}}
\caption{Area of the Shapley Supercluster covered by the present study. 
The clusters of galaxies A3562, A3558 
and A3556 are shown as solid circles the radius of which correspond to 1 Abell radius 
($\sim 3 h_{50}^{-1}$ Mpc). The area of pointed ROSAT PSPC observations 
is indicated by line-shaded circles of $2^{\circ}$ diameter each and labeled by the
ROSAT observation sequence numbers. The region outside 
the pointings is covered by ROSAT All-Sky Survey data.}
\label{fig_region}
\end{figure*}

Fig. \ref{fig_region} shows the field of our study. The clusters A3562, A3558 
and A3556 are shown as solid circles the radius of which correspond to 1 Abell radius 
($\sim 3 h_{50}^{-1}$ Mpc). The ROSAT pointed PSPC observations covering the region
are indicated by line-shaded circles of $2^{\circ}$ diameter each. Table \ref{tab_pointings}
lists the center and total observation time for each observation.

The RASS and pointed PSPC data has been processed following the standard procedure for 
extended X-ray sources. In order to obtain the best signal-to-noise-ratio, only the 
0.5 - 2 keV energy band (the hard ROSAT band) is used (Snowden et al. 1994). This reduces both 
the foreground X-ray emission of the Galactic interstellar medium and the detector background. 
The absorption of the X-ray emission due to neutral hydrogen in the line of sight 
varies only little. The neutral hydrogen column density ranges form
$3.63 \times 10^{20}$ cm$^{-2}$ towards A3558 to $3.91 \times 10^{20}$ cm$^{-2}$ (A3562)
and  $4.05 \times 10^{20}$ cm$^{-2}$ towards A3556 (Stark et al. 1992). 
We adopted in the following the mean value $3.86 \times 10^{20}$ cm$^{-2}$.

\section{X-ray Emission}
\subsection{Spatial analysis}
\begin{figure*}
\centering\leavevmode{\epsfxsize=15.0cm\epsffile{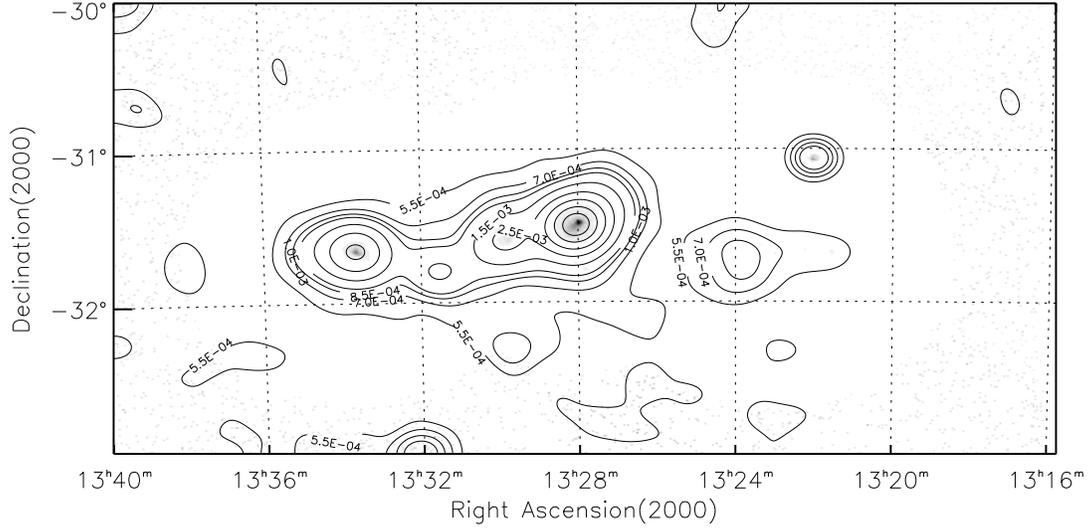}}
\caption{Contours of the X-ray flux of the core region of the Shapley Supercluster. 
The figure shows the count rate in the 0.5 - 2keV energy band (filtered with a Gaussian 
of FWMH=12') obtained from pointed ROSAT PSPC observations and the 
ROSAT All-sky Survey data. The contour levels are 5.5,7,8.5,10,15,25,40,60 and 
$80\times10^{-4}$ cts sec$^{-1}$ arcmin$^{-2}$. X-ray emission tracing A3562 and A3558 
as well as the two groups SC1329-313 and SC1327-312 is clearly visible. On the right 
(i.e. to the west) X-ray emission associated with A3556 is seen.
The contour plot is overlaid on the binned flux. Due to the shorter exposure 
time, the region outside the pointed observations (i.e. the ROSAT All-Sky Survey data) 
exhibits a much lower signal-to-noise ratio resulting in a patchy appearance. Note
the overall elongated structure of the X-ray emission covering 
$\sim 3.75^\circ \times$ 1$^\circ$.}
\label{fig_contour}
\end{figure*}

We began our analysis by merging the data of the pointed ROSAT PSPC observations. We then
binned both the RASS and the pointed data into bins of size 36$\times$36 arcsec$^2$ each
and calculated the corresponding exposure maps using EXSAS. Fig. \ref{fig_contour} 
presents the contour plot of the merged count rates 
in the ROSAT hard band (ROSAT band B, channels 52-201, $\sim$ 0.5-2 keV) obtained 
from the RASS and the pointed data. The contour plot is based on data smoothed with 
a Gaussian of FWMH 12 arcmin. This choice compensates for the different FWMH of the
point spread functions (PSF) of pointed and RASS observations as well as for the varying 
PSF within the aperture of single pointings. The contour plot is overlaid 
on the binned flux. Due to the shorter exposure time, the region outside the pointed 
observations (i.e. the RASS data) exhibits a much lower signal-to-noise ratio resulting 
in a patchy appearance. The figure shows an overall picture of the region. Neither 
the background nor point sources have been subtracted.

Local maxima of the count rate clearly trace the clusters A3562, A3558 and A3556 
as well as the two groups  SC1327-312 and SC1329-313. 
Note also the region in the vicinity of A3556. The peak of the count rate
centered on A3556 is surrounded by faint X-ray emission with an extension of about 
$\sim 1^\circ$ in the east-west direction and $\sim 0.25^\circ$ in the north-south direction.
We have checked for X-ray sources in this region. Clearly visible is the point source 
1RXS J132129.4-314054 of the ROSAT All-Sky Survey bright source catalogue.
Its position roughly coincidents with the western edge of the emission around A3556. 
There are about five other point-like sources in this field, presumably associated with emission
of single galaxies (c.f. section \ref{sec_discuss}). In analogy with a more comprehensive 
study of the X-ray emission of A3558 (Bardelli et al. 1996), we find the point-like 
sources to contribute $< 10\%$ to the observed count rate.

\begin{figure*}
\centering\leavevmode{\epsfxsize=15.0cm\epsffile{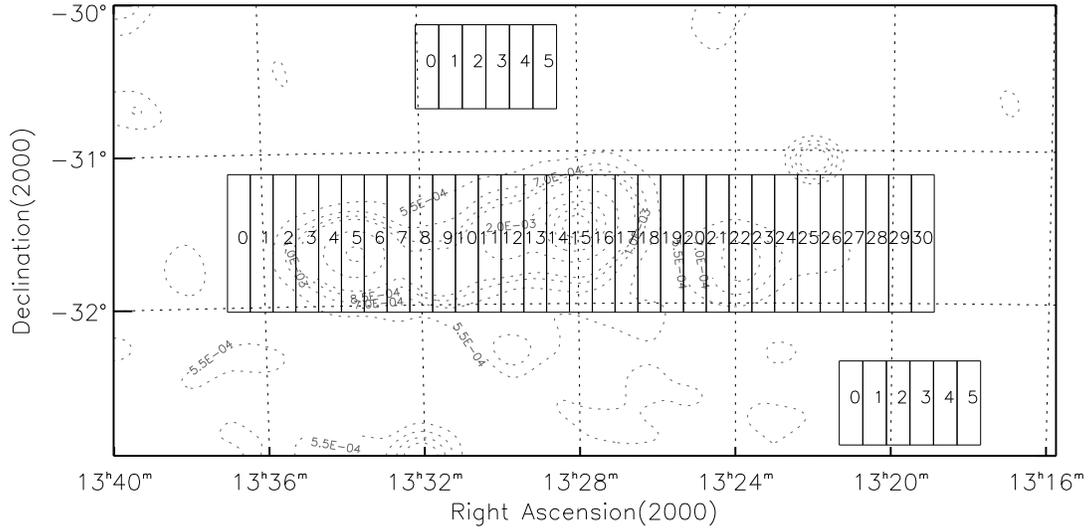}}
\caption{Binning applied in order to check the significance of the elongated X-ray emission.
The count rate in the various bins is plotted in Fig. \ref{fig_flux}. The northern and 
south-western fields have been used for the determination of the background.}
\label{fig_bins}
\end{figure*}

\begin{figure*}
\centering\leavevmode{\epsfxsize=15.0cm\epsffile{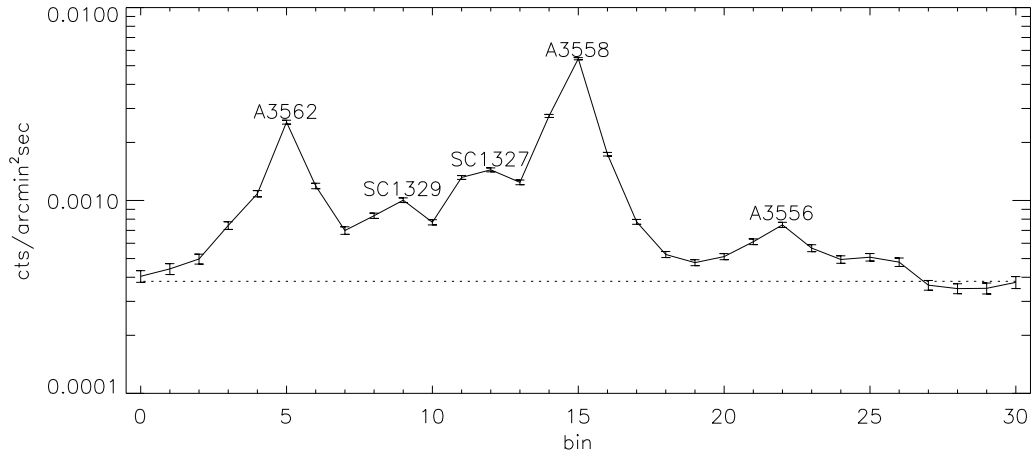}}
\caption{Count rates (in the 0.5-2.0 keV energy band) obtained by binning the flux 
as shown in Fig. \ref{fig_bins}. The error bars 
are the $3\sigma$ errors. The background, taken from the six bins with lowest flux, 
is shown as dotted line. 
The peaks of the count rate are clearly associated with the clusters and groups in the field.
Note that the emission between the clusters A3556 and A3558 lies well above the background.}
\label{fig_flux}
\end{figure*}

To check the significance of the elongated emission, we rebinned the central 
region into 31 bins as shown in Fig. \ref{fig_bins}. The count rates in the various 
bins as obtained from the unfiltered data is plotted in Fig. \ref{fig_flux}. 
The error bars are the $3\sigma$ (Poisson) errors. The background shown as dotted line 
($3.8 \pm 0.35\times10^{-4}$ cts sec$^{-1}$ arcmin$^{-2}$) is estimated from 
the average of the six bins with lowest flux. Its value is in good agreement with 
the background estimated from the smaller northern and south-western fields 
($3.6 \pm 0.57\times10^{-4}$ and $3.7 \pm 0.40\times10^{-4}$ cts sec$^{-1}$ 
arcmin$^{-2}$, respectively). Since the emission found for the background bins shows
some structure, we quote here conservative Gaussian errors. While the 
region used for the first background estimation is covered 
by pointed observations, the two other fields lie mostly within the area of the 
RASS. Because of the poor photon statistics, the data from these two fields has not been 
included in the subsequent analysis.  

The peaks of the flux seen in Fig. \ref{fig_bins} are clearly associated with 
the clusters and rich groups in the field. From Fig. \ref{fig_flux} it is also 
seen that the emission between the clusters A3556 and A3558
lies well above the background. Thus the X-ray emission traces the
the core of the SSC as a filamentary-like structure with an extension of 
$\sim 3.75^\circ$ corresponding to $\sim 17.5$ $h_{50}^{-1}$ Mpc. 

As a final test we checked for the effect of different bin sizes and 
applied the same analysis to the ROSAT A band 
(channels 11-41, $\sim$ 0.1-0.4 keV), C band (channels 52-90 $\sim$ 0.5-0.9 keV) 
and D band (channels 91-201 $\sim$ 0.9-2 keV). While no corresponding emission could be 
found in the A band, the emission observed in the C and D band traces the same 
structure. Therefore we can exclude any contamination by galactic foreground emission 
miming the elongated emission.

Table \ref{tab_lum} shows background subtracted fluxes and luminosities for various 
combinations of bins. For the conversion of the count rates in the ROSAT 0.5-2.0 keV 
band to the flux and luminosity in the 0.1-2.4 keV band we used a Raymond-Smith 
code with metallicity 0.35 (solar units). Where available, the plasma temperature for the 
different regions has been taken from the literature. The plasma temperature of SC1327-312
is estimated to be equal to SC1329-313. For the faint emission between 
the clusters A3558 and A3556 as well as for the elongated emission in the vicinity of
A3556 we assumed a plasma temperature of 1 keV, corresponding roughly to the temperature 
expected for (poor) groups of galaxies or for faint filamentary emission 
(c.f. also section \ref{sec_discuss}). 

The (background subtracted) flux and luminosity obtained for A3558 
(i.e. bins 13-17) are in good agreement
with the corresponding values $F_x=8.4 \pm 0.4 \times 10^{-11}$ erg cm$^{-2}$ s$^{-1}$ and 
$L_x=8.4 \pm 0.1 \times 10^{44}$ $h_{50}^{-2}$ erg s$^{-1}$ (rescaled to the energy range
0.1-2.4 keV) published by Bardelli et al. (1996). Note however, that the 
values given in Table \ref{tab_lum} are of limited relevance since the binning used 
only crudely reflects the actual spatial structure of the different sources. In analogy, 
the total flux and luminosity found for the whole structure by summing up the respective
values for the bins 3-27 has to be considered as a lower limit.

The combination of flux and temperature data with a spatial model for the emitting region 
allows for an estimation of the electron density and gas mass in the respective 
region. In order to get upper limits, we assume here the emitting gas to be 
uniformly distributed in a cylinder whose projection corresponds to the binned region
(i.e. the axis of the cylinder is perpendicular to the line of sight).  
The volume of a bin then is $\pi \times 2.1^2 \times 0.7 \sim 9.7$ $h^{-3}_{50}$ Mpc$^3$.
The resulting electron density and gas mass estimations are quoted 
in Table \ref{tab_lum}. Note that the approximation we applied
overestimates the actual values since it assumes uniform distributions of the flux
and emitting gas. Thus the quoted values have to be considered as upper limits.
The value $M_{gas} \sim 3.8 \times 10^{14} M_\odot$ we find for A3558 (i.e. bins
13-17) is in rough agreement with $M_{gas}=2.5 \times 10^{14} M_\odot$
within $R=2$ Mpc calculated by Bardelli et al. 
(1996). Extrapolating to the whole structure, we find as an 
estimation of the total gas mass $\sim 9 
\times 10^{14} M_\odot$. An more precise determination of the gas mass as well as
the estimation of the total mass should involve a more accurate model of both the gas 
and the flux distribution.

\subsection{Spectral analysis}
Based on ROSAT data, Bardelli et al. (1998) performed a comprehensive analysis of spectral 
properties of the emission of the central cluster A3558 (for a comparison with ASCA data
see Markevitch \& Vikhlinin 1997). Because of comparable exposure times of A3558
and the rest of the region covered by pointed PSPC observations, a corresponding analysis of
the whole elongated emission should be possible. However this is beyond the scope of this 
letter. 
As a first step towards a spectral analysis of the region we determine here the hardness ratio
(i.e. X-ray color) related to the ROSAT C and D energy band
\begin{equation}
h_{\mbox{\tiny C,D}}=
{f_{\mbox{\tiny D}}-f_{\mbox{\tiny C}} \over{f_{\mbox{\tiny D}}+f_{\mbox{\tiny C}}}}
\end{equation}
where $f_{\mbox{\tiny D}}$ and $f_{\mbox{\tiny C}}$ denote the flux in the ROSAT bands C and
D, respectively.

\begin{figure*}
\centering\leavevmode{\epsfxsize=15.0cm\epsffile{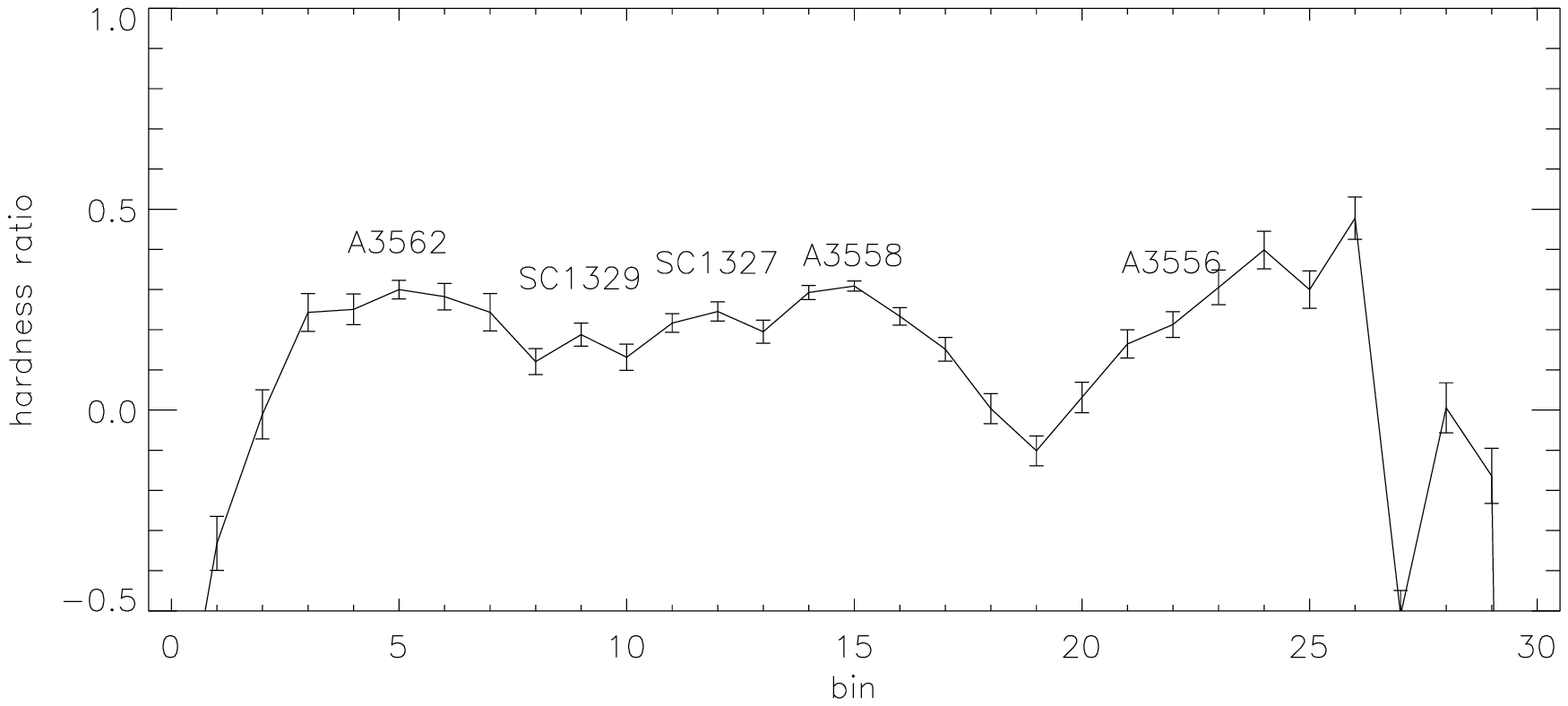}}
\caption{The hardness ratio $[C-D]/[C+D]$ of the background corrected flux 
in the ROSAT C and D band shown as a function of the bin number. The errors are
the $3\sigma$ errors.}
\label{fig_hardness1}
\end{figure*}

In Fig. \ref{fig_hardness1} the hardness ratio of the background corrected flux 
is shown as a function of the bin number. The errors are the $3\sigma$ errors. As it is the 
case for the flux, the hardness ratio traces the structure of the core of Shapely. A clearly 
higher hardness ratio is found for the emission of clusters and groups than for intermediate,
connecting emission. This is evidence for higher gas temperatures associated 
with density peaks, i.e. a result one would expect. For temperatures between 2 and 10 keV,
and a neutral hydrogen column density of $3.86 \times 10^{20}$ cm$^{-2}$ the theoretical 
values for the hardness ratio range form 0.33 to 0.355. These values are
in rough agreement with the hardness ratio found for the clusters and groups in the field.
The corresponding hardness ratio of the background (i.e. the hardness ratio of six bins 
with lowest flux) is -0.22.

For a gas temperature of 1 keV, the theoretical value for the hardness ratio is $0.152$.
As is clear form Fig. \ref{fig_hardness1}, the hardness ratio
between A3556 and A3558 lies bellow this value, indicating a temperature bellow 
1 keV. The oscillating behavior of the hardness ratio westwards of A3556 (i.e. for bins $\ge 25$)
is due to the background subtraction procedure. By subtracting the average flux of 
the six bins with lowest flux, the background corrected net flux in some of these bins 
may become negative giving rise to observed oscillatory feature.

\section{Discussion\label{sec_discuss}} 
The analysis of the ROSAT RASS and pointed data of the core of the SSC
shows clear evidence for X-ray emission related to 
a filamentary superstructure. 
Optical observations (Bardelli et al. 1994, 1998) reveal a striking 
similarity of isodensity contour lines of galaxy counts and X-ray contours. This
strongly suggests a tight connection between the galaxy distribution and
the X-ray emission. It is intriguing that galaxy counts and X-ray emission 
not only correlate on the scale of the single clusters in the field but also on the scale of
the whole filamentary structure, including the region of faint emission between
A3558 and A3556 and westwards of A3556. 

The angular and redshift distribution 
of galaxies in this field provides strong evidence for a filamentary 
superstructure connecting the clusters A3562, A3558 and A3556 (Bardelli et al. 1994). 
We thus conclude the elongated X-ray emission of the core of the SSC 
to trace this filamentary superstructure.

In principle the elongated X-ray emission could arise from projection
effects. However, there is little doubt that this isn't the case. First of all, redshift
measurements indicate the structure to lie perpendicular to the line of sight which
excludes projection effects. Additional evidence comes from the flat, elongated 
X-ray emission of A3556 which is not comparable to the X-ray emission of compact,
isolated clusters.

It is interesting to note that there is additional independent evidence for an 
elongated gas distribution westwards of A3556. The radio survey of 
A3556 (Venturi et al. 1997)
reveals a wide-angle tailed (WAT) radio galaxy at a distance of 
$\sim 2 h_{50}^{-1}$ Mpc westwards from the center of A3556. Normally, WAT 
galaxies are found at the center of groups or clusters and it is assumed 
that the tails are bent by gas flows due to a merging processes 
(Gomez et al. 1997). The original suggestion of Venturi et al. (1997) that 
the bent morphology of the WAT galaxy is due to interaction with surrounding
relatively cold gas (i.e. $T_x < 1$ keV) is confirmed by the present study.   

X-ray emission from filamentary large scale structures is predicted by models (e.g. Bond 
et al. 1996) and recent N-body/hydro simulations (e.g. Cen \& Ostriker 1996) of structure 
formation. It is expected to originate from the hot phase of the intergalactic 
medium (IGM) of temperature $\le 1$ keV, either found in association with
filaments themselves or with single cD galaxies or groups of galaxies
tracing the structure. There have been only few claims of detection of 
X-ray emission on scales extending the scale of clusters of galaxies (i.e. Wang et al. 
1997, Soltan et al. 1997, but see also Briel \& Henry 1995). The fact that 
the present study found clear evidence for X-ray emission on the scale of 
$\sim 17.5 h_{50}^{-1}$ Mpc is reconciled by the extraordinary 
dense nature of the core of Shapley.

While the emission centered on the clusters A3558, A3562 and A3556 as well as 
the one centered on the poor groups SC1329-313 and SC1327-312 is due to hot 
intercluster medium, the nature of the fainter emission connecting the 
whole structure is less obvious. To some extent, it remains debatable if 
the X-ray emission connecting the clusters 
A3558 and A3556 originates from poor groups of galaxies, from IGM 
distributed between the clusters or from overlapping gas distributions. 
In general, the X-ray luminosity of groups of galaxies 
in the 0.1-2.4 keV band is typically $\le 5\times 10^{42}$ erg s$^{-1}$
(Henry et al. 1995, Mulchaey et al. 1996). Comparing this value to the X-ray 
luminosity found in the bins 18-20 leads to $\ge 5$ groups in 
this region, a number which seems rather high. In addition, at least compact 
groups should be resolved as individual sources. We thus exclude this 
possibility.

Likewise, the background subtracted X-ray surface brightness of the bins 18-20 is 
$\sim 2 \times 10^{-15}$ erg cm$^{-2}$ s$^{-1}$ arcmin$^{-2}$ (in the 
0.1-2.4 keV energy band) corresponding to about 2.5 times the upper limit 
for emission from filamentary large scale structure (Briel and Henry, 1995). 
While the extraordinary dense nature of the region could explain this difference,
the overall elongated structure may also suggest that the emission seen in the 
bins 18-20 is due to overlapping gas distributions of A3558 and A3556. In this 
case, the low hardness ratio of the intermediate region could be due to a 
temperature decrease in the outskirts of the clusters. Yet another possibility 
is that we are observing the beginning of the merging of A3558 and A3556.

A more definite answer to this question should involve the determination of the
temperature distribution. If the low temperature (i.e. $\le 1$ keV) assumed 
for the gas in the bins 18-20 is confirmed, this would be 
a strong indication for X-ray emission originating from IGM distributed on
intrasupercluster scale.
The low hardness ratio obtained for the questionable bins as well as the fact 
that we are dealing with an overall exceptional dense region indicates that 
this actually may be the case.

\section{Conclusions}
In summary, our analysis shows that the central part of the Shapley 
Supercluster is traced by elongated X-ray emission connecting the three
clusters of galaxies A3562, A3558, A3556 and the two groups SC1327, SC1329. 
The filamentary-like X-ray emission extends over $\sim 3.75^\circ$ corresponding to 
$\sim 17.5$ $h_{50}^{-1}$ Mpc. This is strong evidence for the original 
claim of Bardelli et al. (1994) that the three clusters A3562, A3558 
and A3556 form a physically connected, single structure.

While the emission between A3558 and A3556 seems not to originate from
groups of galaxies it remains debatable if it is due to overlapping gas 
distributions of A3558 and A3556 or if its of intrasupercluster origin.
A more definite answer to this question as well as to questions related
to the dynamical state of the structure could be provided by the 
temperature distribution of the region. Its determination is thus of great
interest.

\begin{acknowledgements}
It is a pleasure to thank Lev Kofman, Pat Henry and Sandro Bardelli for valuable 
discussions. Part of the work of A.K. has been supported by the Swiss National 
Foundation, grant 81AN-052101. 
\end{acknowledgements}

\newpage

\begin{table*}
\caption{ROSAT Pointed PSPC Observations}
\label{tab_pointings}
\[ \begin{array}
{p{0.2\linewidth} c c c c}
\hline \noalign{\smallskip}
Observation  & \alpha_{2000}  &  \delta_{2000}  & t_{obs} \\
  & \mbox{hh mm ss}& \mbox{dd mm ss} & \mbox{sec}  \\
\hline \noalign{\smallskip}
  800076p   & 13\; 27\; 55 \quad & -31\; 29\; 24 \quad & 29490 \\
  800237p   & 13\; 33\; 38 \quad & -31\; 40\; 12 \quad & 20202 \\
  800375p   & 13\; 22\; 57 \quad & -31\; 43\; 12 \quad & 13763 \\
  800416p-1 & 13\; 23\; 19 \quad & -31\; 30\; 36 \quad & 14598 \\
\hline
\end{array} \]
\end{table*}

\begin{table*}
\caption{Estimated X-ray Flux, X-ray Luminosity, Electron Density and Mass}
\label{tab_lum}
\[ \begin{array}
{p{0.07\linewidth} p{0.07\linewidth} c c c c c c c}
\hline \noalign{\smallskip}
Object  & \mbox{Region} 	
	     & \mbox{Count Rate}^{\rm a} 
	     & T_x  
	     & F_x^{\rm b}
	     & h_{50}^{-2} L_x^{\rm b}
	     & n_e
	     & <M_{gas}\\
   & \mbox{[bins]} 	
	     & \mbox{[cts s$^{-1}$]}   
	     & \mbox{[keV]}
             & \mbox{[erg cm$^{-2}$ s$^{-1}$]}
             & \mbox{[$10^{44}$ erg s$^{-1}$]} 
	     & \mbox{[cm$^{-3}$]}
	     & \mbox{[$10^{14} M_\odot$]}\\
\hline \noalign{\smallskip}
A3562 &   [3-7] & 1.73  & 3.8^{\rm c} & 3.6\times10^{-11} &  3.70 & 1.33 & 2.5\\
SC1329&   [8-9] &  0.43  & 3.0^{\rm c} & 9.0\times10^{-12} &  0.91 & 1.48 & 0.8\\
SC1327&   [11-12] & 0.80  & 3.0^{\rm d} & 1.7\times10^{-11} &  1.70 & 2.01 & 1.1\\
A3558 &   [13-17] & 3.98  & 3.8^{\rm c} & 8.3\times10^{-11} &  8.51 & 2.82 & 3.8\\
      &   [18-20] & 0.15  & 1.0^{\rm d} & 2.4\times10^{-12} &  0.24 & 0.52 & 0.4\\
A3556 &   [21-23]  & 0.31  & 2.1^{\rm c} & 6.3\times10^{-12} &  0.64 & 1.02 & 0.8\\
      &   [24-26]  & 0.13  & 1.0^{\rm d} & 2.2\times10^{-12} &  0.22 & 0.50 & 0.4\\
      &   [3-27]  & 7.53  &  -          & 1.6\times10^{-10} & 15.95 &  -   & 9.8\\
\hline
\end{array} \]
\begin{list}{}{}
\item[$^{\rm a}$] refers to the energy band [0.5-2.0 keV] 
\item[$^{\rm b}$] refers to the energy band [0.1-2.4 keV] 
\item[$^{\rm c}$] $T_x$ is taken from the compilation of White et al. (1997)
\item[$^{\rm d}$] estimation for $T_x$ (see text)
\end{list}
\end{table*}

\newpage

\begin{thebibliography}{}
\bibitem{} Bardelli S., Zucca E., Vettolani G., Zamorani G., Scarmella R.,
Scaramella R., Collins C. A., MacGillivray T., 1994, MNRAS, 267, 665
\bibitem{} Bardelli S., Zucca E., Malizia A., Zamorani G., Scarmella R.,
Vettolani G. 1996, A\&A, 305, 435
\bibitem{} Bardelli, S., Zucca, E., Zamorani, G., Vettolani, G., \& Scaramella, R. 1998, 
MNRAS, 296, 599 
\bibitem{} Bond J. R., Kofman L., Pogosyan D., 1996, Nature, 380, 603
\bibitem{} Breen J., Raychaudhury S., Forman W., Jones C., 1994, ApJ, 424, 59
\bibitem{} Briel U. G., Henry J. P. 1995, A\&A, 302, L9
\bibitem{} Cen R., Ostriker J. P., 1996, ApJ, 464, 27O 
\bibitem{} Dickey J. M., Lockman F. J., 1990, ARA\&A, 28 ,215 
\bibitem{} Drinkwater, M.J., Parker, Q. A., Proust, D., Quintana, H., Slezak, E. 1998,
{\tt astro-ph 9807118}
\bibitem{} Day C. S. R., Fabian A. C., Edge A. C., Raychaudhury S., 1991, 252, 394
\bibitem{} Fabian A. C., 1991, MNRAS, 253, 29
\bibitem{} Gomez P. L., Pinkney J., Burns J. O., Wang Q., Owen F. N., \& 
Voges W. 1997, ApJ, 474, 580 
\bibitem{} Henry J. P., et al. 1995, ApJ, 449, 422
\bibitem{} Lynden-Bell D., Faber S. M., Burstein D., Davies R. L., Dressler A., 
Terlevich R. J., \& Wegner G. 1988, ApJ, 326, 19 
\bibitem{} Ettori S., Fabian A. C., White D. A. 1997, MNRAS, 289, 787 
\bibitem{} Markevitch M., Vikhlinin A. 1997, ApJ, 474, 84 
\bibitem{} Mulchaey J. S., Davis D. S., Mushotzky R. F., Burstein D. 1996, ApJ, 456, 80
\bibitem{} Postman M., Lauer T. R., 1995, ApJ 440, 28 
\bibitem{} Quintana H., Melnick J., Proust D., Infante L., 1997, A\&AS, 125, 247
\bibitem{} Raychaudhury S., Fabian A. C., Edge A. C., Jones, C., Forman W., 1991, MNRAS, 248, 101
\bibitem{} Shapley H., 1930, Harvard Obs. Bull. 874, 9
\bibitem{} Snowden S. C., McGammon D., Burrows D. N., Mendenhall J. A., 1994, ApJ, 424, 714   
\bibitem{} Scaramella R., Baiesi-Pillastrini G., Chincarini G., Vettolani G., Zamorani G.,
1989, Natur, 338, 562
\bibitem{} Soltan, A. M., Hasinger, G., Egger, R., Snowden, S., \& Tr\"umper, J., 
1997, A\&A, 320, 705 
\bibitem{} Tr\"umper, J. 1993, Science, 260, 1769 
\bibitem{} Venturi, T., Bardelli, S., Morganti, R., \& Hunstead, R. W. 1997, MNRAS, 285, 898
\bibitem{} Vettolani G., Chingarini G., Scaramella R., Zamorani G., 1990, AJ 99, 1709
\bibitem{} Voges, W., et al. 1996, MPE Report 263, 637
\bibitem{} Wang Q. D.,  Connolly A. J., Brunner, R. J. 1997, ApJ, 487, L13
\bibitem{} White D. A., Jones C., Forman W. 1997, MNRAS, 292, 419
\end{thebibliography}
\end{document}